\documentclass[aps,prl,twocolumn,showpacs,amsmath,amssymb,superscriptaddress,flushbottom,noraggedfooter]{revtex4-1}

\usepackage{graphicx}
\usepackage{subfigure}
\usepackage{bm}
\usepackage{amssymb}
\usepackage[usenames]{color}

\newcommand{\ket}[1]{\mbox{\ensuremath{| #1 \rangle}}}
\newcommand{\Rb}{$^{87}$Rb}
\newcommand{\figref}[1]{Fig.~\ref{#1}}

\renewcommand\Im{\operatorname{Im}}

\begin{document}

\title{Long-lived periodic revivals of coherence in an interacting Bose-Einstein condensate}

\author{M.~Egorov}
\affiliation{
ARC Centre of Excellence for Quantum-Atom Optics and Centre for Atom Optics and Ultrafast Spectroscopy,\\
Swinburne University of Technology, Melbourne 3122, Australia
}
\author{R.~P.~Anderson}
\affiliation{
ARC Centre of Excellence for Quantum-Atom Optics and Centre for Atom Optics and Ultrafast Spectroscopy,\\
Swinburne University of Technology, Melbourne 3122, Australia
}
\affiliation{School of Physics, Monash University, Victoria 3800, Australia}
\author{V.~Ivannikov}
\altaffiliation{On leave from: Saint Petersburg State University, Russia}
\author{B.~Opanchuk}
\author{P.~Drummond}
\author{B.~V.~Hall}
\author{A.~I.~Sidorov}
\email{asidorov@swin.edu.au}
\affiliation{
ARC Centre of Excellence for Quantum-Atom Optics and Centre for Atom Optics and Ultrafast Spectroscopy,\\
Swinburne University of Technology, Melbourne 3122, Australia
}

\begin{abstract}
We observe the coherence of an interacting two-component Bose-Einstein condensate (BEC) surviving for seconds in a trapped Ramsey interferometer.
Mean-field driven collective oscillations of two components lead to periodic dephasing and rephasing of condensate wave functions
with a slow decay of the interference fringe visibility.
We apply spin echo synchronous with the self-rephasing of the condensate to reduce the influence of state-dependent atom losses, significantly enhancing the visibility up to $0.75$ at the evolution time of $1.5$~s.
Mean-field theory consistently predicts higher visibility than experimentally observed values.
We quantify the effects of classical and quantum noise and infer a coherence time of $2.8$~s for a trapped condensate of $5.5\times10^4$ interacting atoms.
\end{abstract}


\date{\today}

\maketitle

Atom interferometry~\cite{Berman1997,Cronin2009} is a powerful method of precision measurements and a long phase accumulation time
is desirable for improving sensitivity.
Decoherence limits the timescale of an interferometric measurement and is of fundamental importance in quantum information processing.
In this regard interparticle interactions can play a detrimental role~\cite{Sinatra2000,Sinatra2009,Schmiedmayer2010} in interferometry with trapped Bose-Einstein condensates~\cite{Hall1998,Schumm2005,Jo2007}.
Nonlinear interactions generate quantum phase diffusion~\cite{Lewenstein96} and mean-field driven dephasing~\cite{Hall1998,Anderson2009} which lead to the loss of interferometric contrast.
Interaction-induced phase uncertainty has limited the coherence time in a multi-path interferometer to $20$~ms~\cite{Gustavsson2010} and
in a double-well interferometer to $0.2$~s~\cite{Jo2007}.
Monitoring the local spin coherence in the center of a two-component Bose-Einstein condensate (BEC) showed a promising decay time of $0.6$~s~\cite{Lewandowski2003}.
However, spatially nonuniform growth of the relative phase across the BEC leads to fast dephasing of the condensate order parameter and decay of the fringe visibility~\cite{Anderson2009}.

Collisional dephasing can be reduced by tuning the $s$-wave scattering length to zero in the vicinity of a Feshbach resonance~\cite{Gustavsson2008}.
Another way to minimise detrimental interaction effects is to use noncondensed atoms with lower atomic density~\cite{Treutlein2004}.
A long coherence time of $58$~s was recently achieved in Ramsey interferometry with trapped cold atoms using rephasing via the identical spin rotation effect (ISRE)~\cite{Deutsch2010}.

\begin{figure}
        \includegraphics[width=\columnwidth]{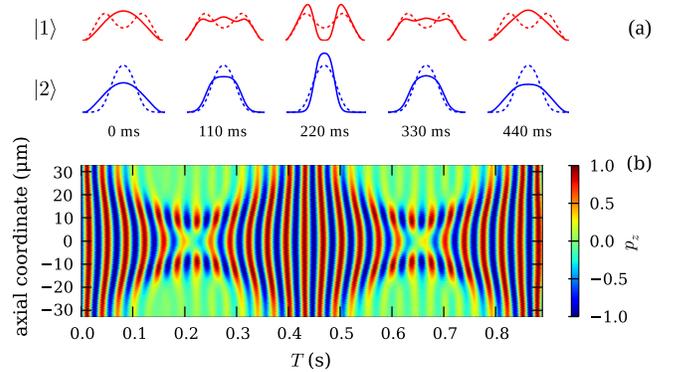}
    \caption[]{(Color) Simulations of collective oscillations of a two-component Bose-Einstein condensate:
        (a) axial density profiles of the components at different evolution times (solid lines) and the two-component ground state (dashed lines);
        (b) normalized local spin projection after Ramsey interferometry (${N=5.5\times10^4}$,~chemical potential $582$~Hz,~${\Delta/2\pi=-37}$~Hz, no losses).
        Lines of constant $p_z$ represent wavefronts of relative phase.
        The mean-field dynamics act to periodically separate the components and curve the phase wavefronts modulating the contrast of interference fringes.
    }
    \label{fig:oscillations}
\end{figure}

In this Rapid Communication we demonstrate that the deleterious effect of atomic interactions on BEC coherence can be reversed via mean-field induced rephasing of the condensate wave functions.
The periodic self-rephasing has a mechanism different from ISRE rephasing observed in noncondensed atoms~\cite{Deutsch2010}
and is due to induced collective oscillations of the condensate wave functions.
The timely application of spin echo further enhances the visibility of Ramsey interference fringes for a trapped \Rb~condensate
and prolongs the coherence time to $2.8$~s.
The pseudospin-$1/2$ system consists of a two-component condensate trapped in two internal states~$\ket{1}$ and $\ket{2}$,
coupled via a two-photon microwave-radiofrequency~(MW-RF) transition.
Four factors affect coherence of the condensate: mean-field dephasing, asymmetric losses, quantum noise, and interaction with noncondensed atoms.
With a single condensate produced in state $\ket{1}$, a $\pi/2$ pulse prepares a nonequilibrium coherent superposition
of two states with modified mean-field interactions and spatial modes distinct from those of the two-component ground state~[\figref{fig:oscillations}(a)].
Small differences in the $s$-wave scattering lengths $a_{11}$, $a_{12}$, and $a_{22}$ cause the wave functions of the two components to oscillate out of phase
around the ground-state modes~\cite{Hall2007}.
During the first half of the oscillation the relative phase grows inhomogeneously~\cite{Anderson2009}, as is evident
from the wavefront curvature of the normalized local spin density $p_z = (n_2 - n_1)/(n_1+n_2)$~[\figref{fig:oscillations}(b)], where
$n_i=\vert\Psi_i\vert^2$ is the density of state $\ket{i}$.
In the second half of the oscillation, the wavefront curvature reverses and the relative phase becomes uniform along the condensate
when the two components complete the cycle and the wave functions overlap again.
The resulting self-revival of the interference contrast would be nearly complete if not for the presence of state-dependent loss of condensate atoms, which we compensate for by using a spin echo technique~\cite{Andersen2003,Oblak2008}.

An \Rb~BEC in the state $\ket{F=1,~m_F=-1}\equiv\ket{1}$ is produced on an atom chip~\cite{Anderson2009} in a cigar-shaped magnetic trap with
radial and axial frequencies $97.0(2)$~Hz and $11.69(3)$~Hz respectively.
The magnetic field at the trap bottom is $3.228(5)$~G for which the first-order differential
Zeeman shift between states $\ket{1}$ and $\ket{2}\equiv\ket{F=2,~m_F=+1}$ is canceled~\cite{Harber2002}.
This ensures that the relative phase is insensitive to magnetic fields and that atomic interactions dominate the dephasing processes.
Radiation for the two-photon coupling is provided by a half-wave dipole antenna
(${f_{\text{MW}}\approx6.8}$~GHz) and on-chip wires (${f_{\text{RF}}\approx 3.2}$~MHz).
The fields are detuned by $1$~MHz from the intermediate state~$\ket{F=2,~m_F=0}$.
The effective two-level system has a two-photon Rabi frequency $\Omega_{12}/(2\pi) = 350$~Hz and
a two-photon detuning $\Delta/(2\pi) = -37$~Hz.
To detect both BEC components simultaneously we use MW adiabatic passage to transfer $99\%$ of state $\ket{1}$ atoms to the $\ket{F=2,m_F=-1}$ state in free fall and apply a magnetic field gradient to spatially separate the components before taking a single absorption image~\cite{Anderson2009}.
In the time-of-flight images we observe no thermal fraction.

\begin{figure}
    \begin{minipage}{0.504\columnwidth}
        \includegraphics[width=\columnwidth]{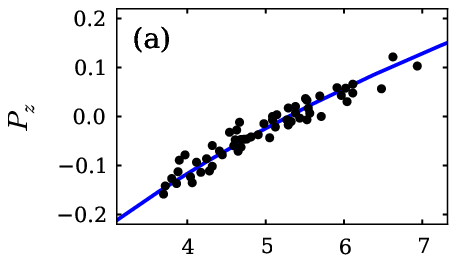} 
        \includegraphics[width=\columnwidth]{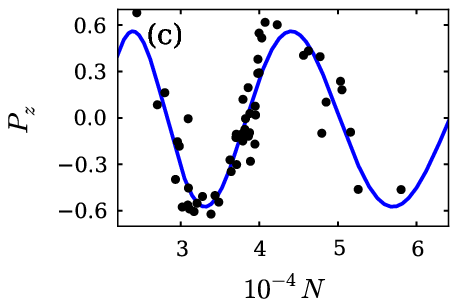} 
    \end{minipage}
    \begin{minipage}{0.456\columnwidth}
        \includegraphics[width=\columnwidth]{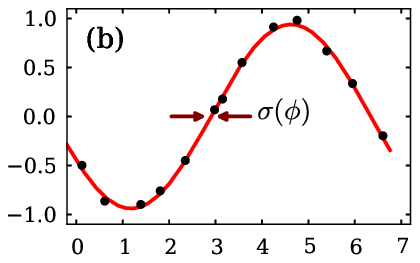}      
        \includegraphics[width=\columnwidth]{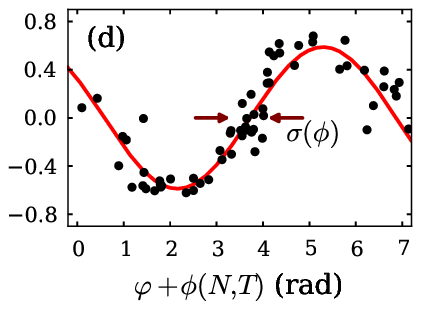}      
    \end{minipage}
    \caption[]{(Color online) Ramsey interference fringes in the atom number~[(a) and (c)] and phase~[(b) and (d)] domains for evolution
        times of~$20$~ms~[(a) and (b)] and $450$~ms~[(b) and (d)].
        Run-to-run variations in the atom number~$N$ cause correlated changes in data points for $P_z$.
        This allows us to correct the phase of the data for the variations of $N$ in phase domain~(b) and (d).
        Solid lines represent expected fringes according to Eq.~(\ref{eq:simplified-pz}).
    }
    \label{fig:fringes}
\end{figure}

The Ramsey sequence starts by applying a $\pi/2$ pulse of $0.7$~ms duration, creating a coherent superposition of $\ket{1}$ and $\ket{2}$.
Both components are confined in the magnetic trap and after an evolution time $T$, an interrogating $\pi/2$ pulse is applied.
The phase lag $\varphi$ of this pulse is varied by a programmable MW source in order to obtain a Ramsey fringe
in the phase domain.
Immediately after the second pulse we release the atoms, measure the populations $N_1$ and $N_2$ from an absorption image and evaluate
the normalized atom number difference ${P_z=(N_2-N_1)/(N_1+N_2)}$.
In the mean-field formalism
\begin{equation}
    \label{eq:gp-pz}
    P_z(N,T,\Delta,\varphi)=\frac{2}{N}\Im\left[e^{i(\varphi + \Delta\, T)}\int \Psi_2^{*}\,\Psi_1\,d^3\mathbf{r}\right],
\end{equation}
where $N$ is the total atom number at $T$.
To analyze the interference fringes we use the expression
\begin{equation}
    \label{eq:simplified-pz}
    P_z(N,T,\Delta,\varphi)=\mathcal{V}(N,T)\cos[\Delta\,T + \phi(N,T) + \varphi],
\end{equation}
where $\phi(N,T)$ is the interaction-induced relative phase, and $\mathcal{V}$ is the fringe visibility.
Fringes in $P_z$ strongly depend on $N$ as shown in~Figs.~\ref{fig:fringes}(a) and~\ref{fig:fringes}(c).
While experimental control over $\Delta$, $T$, and $\varphi$ is straightforward, run-to-run fluctuations in the total number of condensed atoms lead to
systematic shifts of the fringe.
The effect of such variations is corrected for by using the measured number of atoms in each experimental cycle.
We approximate $\phi(N,T) = \alpha(T)\,N^{2/5}\,T$, where $\alpha(T)$ is determined from simulations of the
Gross-Pitaevskii equations~(GPE) and slowly varies with $T$; $\alpha(20~\text{ms})=0.79$ and $\alpha(450~\text{ms})=0.90$.
This approximation for $\phi(N,T)$ agrees well with the fringes acquired by varying the atom number with constant phase $\varphi$~[Figs.~\ref{fig:fringes}(a) and ~\ref{fig:fringes}(c)].
For each $T$ we vary the microwave phase $\varphi$ and obtain Ramsey fringes in the phase domain~$\varphi+\phi(N,T)$, corrected for number fluctuations~[Figs.~\ref{fig:fringes}(b) and~\ref{fig:fringes}(d)],
from which we evaluate the interference contrast $\mathcal{V}$.

In recent spin-squeezing experiments~\cite{Esteve2008,Riedel2010,Oberthaler2010} the atom number measurement was calibrated using the scaling
between $N$ and the quantum projection noise.
We calibrate the atom number measurement
using the simulated dependence of $P_z$ on the total number $N$ at the evolution time of $340$~ms~(for $500<N<10^5$).
We numerically solve the three-dimensional coupled GPE equations~\cite{Hall2007,scattering-lengths}
to obtain the wave functions $\Psi_1$ and $\Psi_2$ and evaluate $P_z$ using Eq.~(\ref{eq:gp-pz}).
The best fit of the model yields an absolute calibration of $N$ with a statistical uncertainty of $5\%$.
\begin{figure}
    \includegraphics{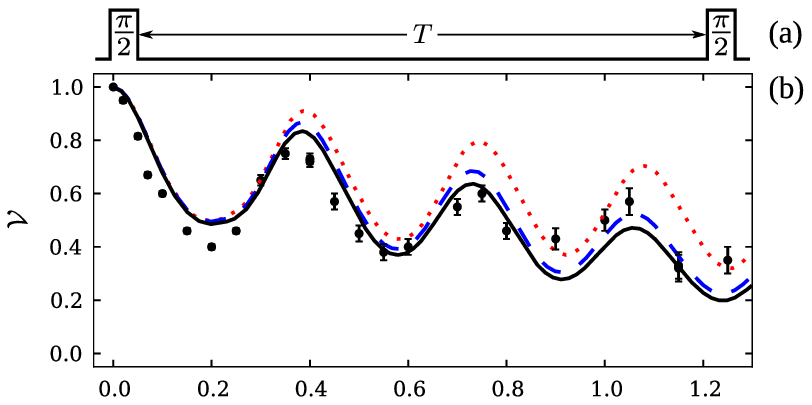}
    \vskip 0.2cm
    \includegraphics{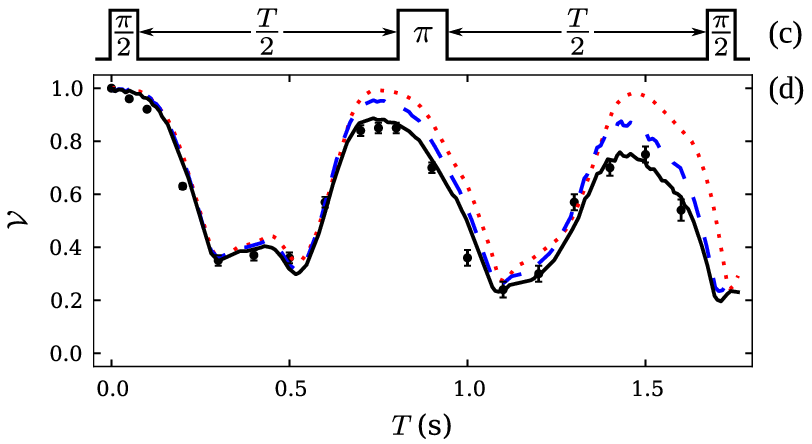}
    \caption{(Color online)
    Schematic of interferometric sequence for (a) Ramsey and (c) spin echo. Corresponding plots of
    the visibility $\mathcal{V}$ [(b), (d)] where data (filled circles) are measured in the phase domain as a function of $T$.
    Curves correspond to simulations using coupled GPE theory (red dotted line), truncated Wigner model with classical noise (blue dashed), and 
    with additional $10$~s exponential decay (solid black).
    }
    \label{fig:visibility}
\end{figure}

In the temporal evolution of the Ramsey fringe visibility $\mathcal{V}(N,T)$ (initial
total atom number $5.5(6)\times 10^4$, peak density $7.4\times10^{13}~\textrm{cm}^{-3}$) we observe a decaying periodic structure with peaks and troughs~[\figref{fig:visibility}(b)].
The initial decrease of visibility is due to the nonuniform growth of the relative phase~\cite{Anderson2009} and the spatial
separation of the components~\cite{Hall2007}.
By the end of the first collective oscillation the relative phase is uniform and the two components overlap again~(\figref{fig:oscillations}).
This periodic evolution continues with a slow decay of fringe visibility (decay time of $1.3$~s).
Inelastic two-body collisions remove atoms from state $\ket{2}$ faster than from state $\ket{1}$~\cite{scattering-lengths} limiting the maximum fringe contrast to
${\mathcal{V}_{\max} = 2\sqrt{N_1 N_2}/(N_1+N_2)}$.
The coupled GPE model [\figref{fig:visibility}(b), dotted line] correctly predicts the period of the visibility revivals ($T_{\text{rev}}=0.37$~s),
but overestimates their magnitudes.

We apply a spin echo pulse~[\figref{fig:visibility}(c)] to reverse the relative phase evolution and compensate for asymmetric losses by
inverting the populations.
Maximum revivals of the visibility are observed ($\mathcal{V}=0.85$ at $T=0.75$~s and $\mathcal{V}=0.75$ at $T=1.5$~s) when the self-rephasing and spin echo work in unison.
This occurs if the $\pi$ pulse is applied at $T/2 = T_{\text{rev}}, 2\,T_{\text{rev}}, \ldots$, when the relative phase is uniform along the condensate.
The GPE simulations predict an almost complete revival of the visibility ($\mathcal{V}=0.98$ at $T=1.5$~s), which is not seen
in the experiments~[\figref{fig:visibility}(d)].

\begin{figure}
    \centering
    \includegraphics{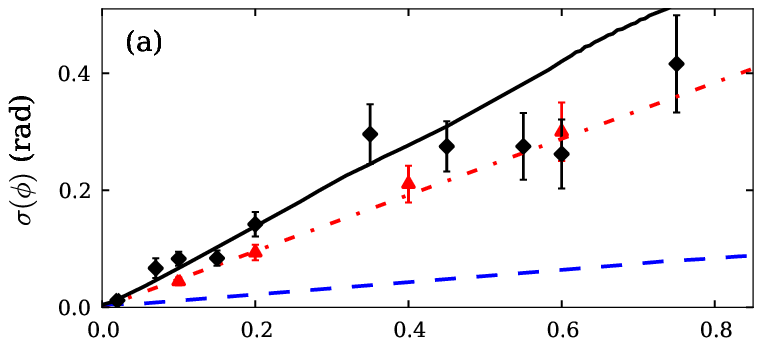}\\
    \includegraphics[width=0.9\columnwidth]{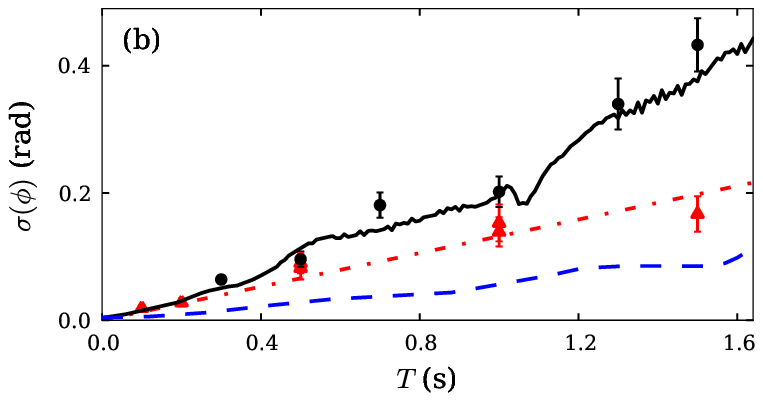}
    \caption{(Color online)
	Phase uncertainty growth in Ramsey (a) and spin echo (b) sequences with the two-component BEC (black circles) and noncondensed cold atoms (red triangles).
	The blue dashed lines show the effect of quantum noise simulated with the truncated Wigner method.
	The red dash-dotted lines indicate the contribution from MW frequency fluctuations.
	The black solid lines show the simulated combined effect of quantum and classical noise.
	}
    \label{fig:phase}
\end{figure}

To characterize phase decoherence we studied the growth of the phase uncertainty for Ramsey and spin echo sequences
(\figref{fig:phase}).
For each time $T$ we set the phase $\varphi$ of the second $\pi/2$ pulse to ensure $\langle P_z\rangle = 0$.
In each realization we measure $N$, record the deviation $\delta P_{z,i}$ from the computed atom number dependence
[Figs.~\ref{fig:fringes}(a) and ~\ref{fig:fringes}(c), solid line] and record the
relative phase deviation $\delta\phi_i = \arcsin\left(\delta P_{z,i}/\mathcal{V}\right)$.
We use $M=50$ to $100$ realizations for each datum point in \figref{fig:phase} and evaluate the Allan deviation $\sigma(\phi)$ for the duration of an experimental cycle (50~s).
Error bars on each point in \figref{fig:phase} represent the statistical uncertainty of $\sigma(\phi)$ and are calculated as $\sigma(\phi)/\sqrt{M-1}$.

We evaluated contributions from three sources of classical noise:
MW frequency fluctuations,
uncertainty in the recorded atom number, and fluctuations of the splitting (spin echo, recombination) process.
To characterize the MW frequency instability we measured growth rates of the phase uncertainty of $0.50(8)$ rad/s in the Ramsey and $0.125(20)$ rad/s in the spin echo sequences (\figref{fig:phase}, dash-dotted lines) for $5.5\times10^4$ uncondensed cold atoms trapped at $160$~nK.
Frequency fluctuations in our imaging laser contribute an uncertainty to the recorded atom number, increasing the spread of $\delta P_{z,i}$ [\figref{fig:fringes}(a)].
The probe laser linewidth of $1.3$~MHz (full width at half maximum, FWHM) results in a $2.3\%$ uncertainty in the measured atom number when imaging on resonance.
We find the contribution of the measured atom number uncertainty to the phase noise by calculating $d\phi/dN$ from the GPE simulations.

We also considered the role of quantum noise in decoherence of the condensate
through quantum phase diffusion~\cite{Lewenstein96} and nonlinear losses.
We use the truncated Wigner approach~\cite{Steel1998,Opanchuk11} and
start from a Born-Markov master equation with additional terms for the dominant two-body losses~\cite{scattering-lengths}.
The master equation is then transformed to a multimode Wigner representation, and the resulting partial differential equation for the Wigner quasiprobability function is truncated, leaving only first- and second-order derivatives.
The equivalent stochastic equations are solved numerically, and
the simulated quantum phase diffusion is plotted in \figref{fig:phase} (dashed lines).
In the experiment, the fluctuations of $P_z$ after splitting are not limited by quantum noise
but are determined by shot-to-shot variations and drift of the two-photon Rabi frequency $\Omega_{12}$.
Over several hours we typically measure a $2\%$ deviation of $\langle P_z\rangle$ after the first $\pi/2$ pulse,
which is above the standard quantum limit of $0.43\%$.
After the inclusion of $2\%$ splitting fluctuations into the simulations we obtain the phase diffusion induced by quantum and splitting noise.
The black solid line~(\figref{fig:phase}) represents the combined contribution from the classical and quantum sources of noise and is very close to the spin echo data.

The GPE simulations of the temporal evolution of visibility~(\figref{fig:visibility}) do not include irreversible decoherence effects and
consistently show higher values than the experimental data.
If phase noise $\sigma(\phi)$ is the major source of decoherence the visibility would decrease by a coherence factor $e^{-\sigma(\phi)^2/2}$~\cite{Schmiedmayer2010,Gati2007}, which is
equal to $0.92$ for $T=0.75$~s in the Ramsey sequence and for $T=1.5$~s in the spin echo sequence; a value which cannot explain
the detected decrease of visibility in both cases.
In addition to the interaction-induced phase diffusion our quantum noise simulations take into account asymmetric two-body losses which are significant since $75\%$ of condensed atoms are lost (predominantly through the state $\ket{2}$ decay) at the spin echo evolution time of $1.5$~s.
The modeling shows that at $T=1.5$~s the mean value of the total pseudospin is $\langle\hat{S}^2\rangle^{1/2}=0.924(N/2)$, an additional factor reducing the interference contrast.
The combined effect of particle losses and quantum and classical phase diffusion on the visibility in the spin echo and Ramsey sequences is shown in~\figref{fig:visibility} (blue dashed line).
In order to reconcile the remaining deviation of experimental points from the simulation results we introduce single-particle decoherence, possibly associated with the interaction of the condensate with uncondensed thermal atoms~\cite{Sinatra2009}.
A previous study~\cite{Lewandowski2003} demonstrated that the condensate presence shortens the coherence time of a thermal cloud,
but the reverse effect of the uncondensed atoms has not been recorded.
In our experiment the thermal fraction is small, as we do not observe its presence at the start or the end ($1.5$~s) of the interferometric sequence.
We introduce an additional exponential decay with a time constant of $10$~s to our simulations for the best fit with the spin echo data (black solid line in~\figref{fig:visibility})
to infer an overall coherence time of $2.8$~s.

In conclusion, we report that atomic interactions not only lead to quantum and spatial dephasing of a trapped BEC interferometer but also produce mean-field-driven rephasing through periodic collective oscillations.
The synchronized application of spin echo significantly enhances the self-rephasing, allowing us to record a visibility of $0.75$ at an evolution time of $1.5$~s.
The quantum noise simulations consider the effects of asymmetric two-body losses and quantum phase diffusion on the BEC coherence.
Together with the evaluations of classical noise and single-particle coherence, the simulations allow us to estimate a condensate coherence time of $2.8$~s,
the longest coherence time for an interacting quantum degenerate gas.
Future work will focus on the suppression of classical noise, reaching longer coherence times in the quantum noise dominated regime and exploring the effect of thermal atoms on condensate decoherence~\cite{Sinatra2009}.
Our observations show that long coherence times are possible in a trapped interacting BEC interferometer, which is an important result for precision measurements and sensor development based on quantum degenerate gases.

We acknowledge helpful discussions with P.~Hannaford and S.~Whitlock.
The project is supported by the ARC Centre of Excellence for Quantum-Atom Optics, and an ARC LIEF grant LE0668398.

\bibliography{rephasing}

\begin{thebibliography}{26}
\bibitem [{sca()}]{scattering-lengths}%
  \BibitemOpen
  \href@noop {} {}\bibinfo {note} {In the simulations we used the scattering
  lengths $a_{11}=100.4\,a_0$~\cite{Hall2007}, $a_{22}=95.57(15)\,a_0$, and
  $a_{12} = 97.993(20)\,a_0$ that we measured from the collective oscillations
  (to be published) and the loss rates
  $\gamma_{12}=1.51(18)\times10^{-14}~\mbox{cm}^3/\mbox{s}$ and
  $\gamma_{22}=8.1(3)\times10^{-14}~\mbox{cm}^3/\mbox{s}$, which were obtained
  from our measurements of atom number decay.}\BibitemShut {Stop}%
\end{thebibliography}

\end{document}